\def\beq{\begin{equation}}
\def\eeq#1{\label{#1}\end{equation}}
\def\eeqn{\end{equation}}

\def\beqa{\begin{eqnarray}}
\def\eeqa#1{\label{#1}\end{eqnarray}}
\def\eeqan{\end{eqnarray}}

\let\bar=\overbar

\def\Dslash{\not{\hbox{\kern-4pt $D$}}}
\def\dslash{\not{\hbox{\kern-2pt $\del$}}}

\def\msb{{\bar{\ssstyle M \kern -1pt S}}}

\documentclass[12pt]{article}
\usepackage{epsfig}

\def\Title#1{\begin{center} {\Large {\bf #1} } \end{center}}

\begin{document}
{\hfill \bf UF-IHEPA 02-01}

\Title{A Review of Charmed Baryon Experimental Data}

\bigskip\bigskip


\begin{raggedright}  

{\it John Yelton\index{Yelton, J.~M.}\\
Physics Department\\
U. of Florida\\
Gainesville, FL 32611-8440}
\bigskip\bigskip
\end{raggedright}

\section{Introduction}

This is a review of the experimental results on charmed baryons, with the 
accent on those reported most recently.

\section{Motivation}

The charmed baryon sector is the richest quark spectroscopy 
available for study. Compared with mesons there are {\it more} states as there
are more possibilities for orbital excitations. More importantly, the extra
mass associated with these excitations is less (because it is inversely 
proportional to a measure of the center-of-mass), leading to less phase space
for decays and {\it narrower} states. The charm quark is sufficiently
massive for the states to be described as a combination of a heavy quark and a
light di-quark - this picture does not work well for strange baryons. 
The charmed baryon sector is much easier to study experimentally than the $B$
baryon sector. It also offers a good laboratory for studying weak decays
as there are four weakly decaying singly charmed baryons.

\section{Techniques}

There are two main techniques that have been used in recent years for charmed
baryons studies. Fixed target experiments, such as FOCUS, SELEX and E-791 at
Fermilab, have a long pathlength for charm decays which can be used as a tag,
and also can yield accurate lifetime measurements. Spectroscopy has tended to be
easier at $e^+e^-$ machines operating at around 10 GeV (notably CLEO, now joined 
by BELLE). In the $e^+e^-$ continuum, around 40\% of events are charm.
Running at the $\Lambda_c^+$ threshold has not been done for a long time but there
is hope that CLEO\_c will do this in the future.

\section{The Greek Alphabet of Charmed Baryons}

We always consider a charmed baryon as the combination of a (heavy) charm quark 
and a light di-quark, which has its own well-defined quantum numbers
$J^P_{LIGHT}$. This combines with the charm quark to give the overall 
$J^P$ of the state. If the two light quarks are $u$ and/or $d$ then the particle is
either a $\Lambda_c$ or a $\Sigma_c$. The former are anti-symmetric under interchange
of the two light quarks and are iso-scalars, the latter are symmetric under 
interchange and are iso-triplets.  When one of the two light quarks is 
a strange quark the baryons are called $\Xi_c$ states, and if both light 
quarks are strange it is an $\Omega_c$.

\section{The $\Lambda_c^+$ Ground State}

More than thirty decay modes of the $\Lambda_c^+$ have been measured. A recent
contribution comes from BELLE\cite{BELLELC} 
who have measured a series of comparatively rare
decays that are either Cabibbo-suppressed (such as the first observation 
of $\Lambda K^+$) or unambiguously due to W-exchange diagrams. 
There have also been two recent measurements of the $\Lambda_c^+$ lifetime, 
from SELEX\cite{SELEXLC}
($\tau(\Lambda_c^+)=198.1\pm7.0\pm5.6$ fs) and FOCUS $(\tau(\Lambda_c^+)=
204.6\pm3.4\pm2.4$ fs \cite{FOCUSLC}). The PDG 2001\cite{PDG2001} 
number of of $188\pm7$ 
seems to be edging up. 
This is of course a short lifetime by charm standards, 
presumably because W-exchange
is an allowed method for $\Lambda_c^+$ decays, as has been shown directly. 

\section{Recent $\Sigma_c$ and $\Sigma_c^*$ Results}

The doubly charged and neutral  $\Sigma_c$ states are 
relatively easy to observe as they decay with a charged pion to the 
$\Lambda_c^+$ ground state. 
The masses of all three $\Sigma_c$ states have been well measured. Two 
recent measurements from CLEO\cite{CLEOSC} and FOCUS\cite{FOCUSSC} 
measure the natural widths, 
with all results being around 2 MeV. 
The singly charged states decay via a
$\pi^0$ decay which is usually harder to detect experimentally.
There are fewer measurements of the
 $\Sigma_c^{*}$ states because their large natural widths cause 
complications. The singly charged state was only recently reported
by CLEO\cite{CLEOSCS}. 
Looking at all the results, we can note 
that a) there is little isospin splitting, but
the singly charged state may be a little lighter than the others
in line with predictions\cite{SCPRED}, 
b) the width of the ${3\over{2}}$ states is around 
seven times that of the ${1\over{2}}$ states - this is in line with
predictions that this ratio depends only on a few simple numerical 
factors plus phase-space\cite{SCROSNER}, 
and 
c) a very naive quark model which predicts that the mass
splitting between the spin-weighted average of the $\Sigma^{*}$-$\Sigma$
system and the ground state should be independent of the heavy quark mass,
and the splitting between the ${3\over{2}}$ and ${1\over{2}}$ states
is inversely proportional to heavy quark mass. 
The strange and charmed
$\Sigma-\Lambda$ system obeys this scaling law very well. 
I expect it to do so in the B-system also.

\section{Higher States}

Allowing orbital angular momentum into the picture produces a large
number of predicted states. Some of these can be expected to be narrow
and some very wide. With the $L=1$ between the heavy quark and the 
light di-quark, there should be 2 $\Lambda_{c1}$ states (well known
and well measured) and no fewer than 5 iso-triplets of $\Sigma_c$ 
states. Most people use a numerical subscript to denote the spin 
of the light diquark. 
Alternatively, having $L=1$ between the two light quarks
produced 5 $\Lambda_c$ states and 2 iso-triplets of $\Sigma_c$'s.
All these particles (except maybe some above $pD$ threshold), will 
cascade down via (multi-)pion decays to the ground-state $\Lambda_c$.
CLEO have found two bumps in $\Lambda_c^+\pi^+\pi^-$\cite{CLEOXY}. 
One is wide,
and they like the identification as the first two $\Sigma_{c1}$ states.
The second is more interesting as it is fairly narrow, and they 
identify it as the first of the second generation of orbitally
excited $\Lambda_c$ particles. This particular state has no allowed
single pion decay available. These observations have yet to pique the
theorists interest. Figure 1 shows a guess of the spectroscopy of 
the $\Lambda_c - \Sigma_c$ states, based upon a very simple potential model. 
The states that CLEO guesses correspond to their bumps are denoted by bold lines.

\section{The $\Xi_c$ Spectrum}

The $csu$ and $csd$ quark combinations are referred to as $\Xi_c^+$ and 
$\Xi_c^0$ respectively. Their spectroscopy follows the lines of the
$\Lambda_c - \Sigma_c$ pattern, except that it comprises only 
iso-doublets, rather than iso-singlet and iso-triplets. The first 10 states
(analagous to the those up to and including the $\Lambda_{c1}(2630)$) have
been reported. Nine of these first observations were by CLEO\cite{CLEOXIC} and many of
them have yet to be confirmed. The $\Xi_c^+$ groundstate has been known 
since the eighties (although first sightings\cite{XIC+} were controversial), and its
lifetime has been measured several times with a PDG (2001)\cite{PDG2001} 
average of
$330^{+60}_{-40}$ fs. Since then have been two more measurements of this lifetime.
The CLEO II.V detector selects a clean sample of $\Xi_c^+$ decays, but its
pathlength resolution is comparable to the individual pathlengths. It finds
$503\pm47\pm18$ fs\cite{CLEOXICL}. FOCUS has more events\cite{FOCUSXICL}, 
worse signal-to-noise, but much better
lifetime resolution. It obtains $439\pm22\pm9$ fs. It seems that the $\Xi_c^+$ 
lifetime is creeping up. 

\section{The $\Omega_c$}

The $\Omega_c (css)$ combination was reported many times over the last two 
decades, but we can safely say now that most have been shown to be
incorrect on mass and/or cross-section grounds. The E-687 peak in 
$\Sigma^+K^-K^-\pi^+$ is the one that still looks impressive\cite{E687}. 
In 2001
CLEO\cite{CLEOOC} found a good looking peak using the sum of 5 ``expected'' decay modes
(not including $\Sigma^+K^-K^-\pi^+$). Now BELLE\cite{BELLEOC} 
has (preliminary) results
shoding an extremely clean peak in $\Omega^-\pi^+$. Their mass 
($2693.7\pm1.3^{+1.1}_{-1.0}\ $ MeV) agrees well with the CLEO number
($2694.6\pm2.6\pm1.9\ $ MeV). There is no doubt that the $\Omega_c$ has
been discovered.

A recent analysis by CLEO finds semi-leptonic decays of the $\Omega_c$.
They reconstruct $760\pm32$ $\Omega^-$ baryons (it is clearly tough to 
produce an $sss$ state in $e^+e^-$ annihilations). They then look for correlations
with correctly charged electrons. After subtraction of 
backgrounds, they find an excess of $11.4\pm3.8$ events
that are attributable to $\Omega_c \to \Omega^- e^+ \nu$ events. 
The ratio of the simplese hadronic decay mode $\Omega^-\pi^+$ to this 
semi-leptonic mode is $0.41\pm0.19\pm0.04$. This is similar to analagous 
ratios in $\Xi_c^0$ and $\Lambda_c^+$ decays. 

\section{Doubly charmed baryons}

Theorists have always enjoyed predicting the masses of the doubly charmed 
baryons $\Xi_{cc}^{++}\ (ccu)$ and $\Xi_cc^+\ (ccd)$. SELEX have shown\cite{SELEXCCU}
peak (of only 3 $\sigma$ significance and definitely
preliminary) in the decay mode $\Lambda_c^+K^-\pi^+\pi^+$. The mass is
3.79 GeV. I do not understand what processes must be involved 
to produce enough of them so that an individual decay mode of one of the 
states will make around 1\% of the $\Lambda_c^+$ candidates in their sample.      
  
\section{Conclusion}

There have been 22 charmed baryon states reported in the literature, though 
many of them need to be confirmed. They display a spectroscopy that is complex, 
yet orderly and comprehensible. There remains more work to be done on 
spectroscopy, and work is still active in understanding decay mechanisms.
In the future, BELLE and BaBar are the experiments best placed for new
discoveries, and CLEO\_c operating at $\Lambda_c^+$ threshold could
also make complementary contributions.

\begin{figure}[htb]
\begin{center}
\epsfig{file=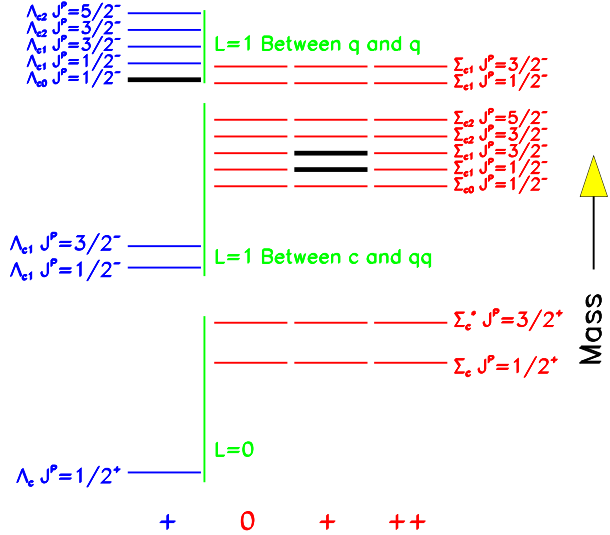,height=3.5in,bbllx=50,bblly=190,bburx=570,bbury=720}
\caption{The author's view of the expected mass spectrum of singly-charmed baryons.}
\label{fig:magnet}
\end{center}
\end{figure}


\begin{thebibliography}{99}


\bibitem{BELLELC}
K.Abe {\it et al.,} Phys. Lett {\bf B524} 247 (2002).

\bibitem{SELEXLC} A.~Kushnirenko {\it et al.,}
PRL {\bf 86} 5243 (2001).

\bibitem{FOCUSLC} J.~M.~Link {\it et al.,} PRL {\bf 88}, 161801
(2002).

\bibitem{PDG2001} D.~E.~Groom {\it et al.,} Eur. Phys. J. {\bf C15}
(2000), 1. with 2001 update (URL:http://pdg.lbl.gov/).

\bibitem{CLEOSC}
M.~Artuso {\it et al.,} Phys. Rev. {\bf D65} 071101, (2002).

\bibitem{FOCUSSC} J.~M.~Link {\it et al.,} Phys. Lett. {\bf B525},
205 (2002).

\bibitem{CLEOSCS} R.~Ammar {\it et al.,} PRL {\bf 86}
1167 (2001).

\bibitem{SCPRED} J.~Franklin, Phys. Rev. {\bf D 59}, 117502
(1999).

\bibitem{SCROSNER} J.Rosner, Phys. Rev. {\bf D 52}, 6461 (1995).


\bibitem{CLEOXY} M.~Artuso {\it et al.,} PRL {\bf 86} 4479 (2001).

\bibitem{CLEOXIC} S.E.Csorna {\it et al.,} PRL {\bf 86}, 2232 (2001),
J.~Alexander {\it et al.,} PRK {\bf 83} 3390 (1999), C.~P.~Jessop 
{\it et al.,}  PRK {\bf 82} 492 (1999), L. Gibbons {\it et al.,}
PRL {\bf 77}, 810 (1996), P.~Avery {\it et al.,} PRK {\bf 75},
4364 (1995),

\bibitem{XIC+} S.~Biagi {\it et al.} Phys. Lett. {\bf 122B} 455 (1983).
 
\bibitem{CLEOXICL}
A.H.Mahmood {\it et al.,} Phys. Rev. {\bf D65} 031102 (2002).

\bibitem{FOCUSXICL} J.~M.~Lin {\it et al.,} Phys. Lett. {\bf B523}
(2001).

\bibitem{E687} P.~L.~Frabetti {\it et al.,} Phys. Lett. {\bf B338}
106 (1994).

\bibitem{CLEOOC} D.Cronin-Hennessy {\it et al.,} 
PRL {\bf 86} 3730 (2001).

\bibitem{BELLEOC} Private communication, P.~Pakhlov (ITEP) on behalf of BELLE.

\bibitem{SELEXCCU} For latest SELEX results, see their homepage
http:///fn781a.fnal.gov/






\end{thebibliography}
\end{document}